%% file: main.tex
\documentclass[conference]{IEEEtran}
\IEEEoverridecommandlockouts

\usepackage{cite}
\usepackage{caption}
\usepackage{amsmath,amssymb,amsfonts}
\usepackage{algorithmic}
\usepackage{graphicx}
\usepackage{siunitx}
\usepackage{textcomp}
\usepackage[dvipsnames]{xcolor}
\usepackage{makecell}
\usepackage{booktabs}
\usepackage[acronym]{glossaries}

\newacronym{cmos}{CMOS}{Complementary Metal-Oxide-Semiconductor}
\newacronym{if}{I\&F}{Integrate-and-Fire}
\newacronym{lif}{LIF}{Leaky Integrate-and-Fire}
\newacronym{aer}{AER}{Address-Event Representation}
\newacronym{snn}{SNN}{Spiking Neural Network}
\newacronym{rf}{R\&F}{Resonate-and-Fire}
\newacronym{vlsi}{VLSI}{Very-Large-Scale Integration}
\newacronym{lv}{LV}{Lotka–Volterra}
\newacronym{req}{REQ}{request}
\newacronym{ack}{ACK}{acknowledge}
\newacronym{cv}{CV}{Coefficient of Variation}
\newacronym{fi}{F-I}{frequency–current}
\newacronym{emg}{EMG}{electromyography}
\newacronym{ecg}{ECG}{electrocardiography}
\newacronym{ai}{AI}{Artificial Intelligence}
\newacronym{qf}{Q-factor}{quality factor}
\newacronym{asic}{ASIC}{Application-Specific Integrated Circuit}

\renewcommand{\baselinestretch}{0.97}

\def\BibTeX{{\rm B\kern-.05em{\sc i\kern-.025em b}\kern-.08em
    T\kern-.1667em\lower.7ex\hbox{E}\kern-.125emX}}
\begin{document}

\title{An Asynchronous Mixed-Signal \\ Resonate-and-Fire Neuron\\

\thanks{This work was supported by the European Research Council (ERC) through the European Union's Horizon Europe Research and Innovation Programme under Grant Agreements No. 101119062 and 101042585. The authors would like to acknowledge the financial support of the CogniGron research center and the Ubbo Emmius Funds (Univ. of Groningen).
}
}

\author{\IEEEauthorblockN{Giuseppe Leo\IEEEauthorrefmark{1},
Paolo Gibertini\IEEEauthorrefmark{1},
Irem Ilter\IEEEauthorrefmark{2},
Erika Covi\IEEEauthorrefmark{1},
Ole Richter\IEEEauthorrefmark{3}\IEEEauthorrefmark{4},
Elisabetta Chicca\IEEEauthorrefmark{1}}
\IEEEauthorrefmark{1}\footnotesize Zernike Institute for Advanced Materials and Groningen Cognitive Systems and Materials Centre (CogniGron),\\
University of Groningen, Groningen, The Netherlands\\
\IEEEauthorrefmark{2}\footnotesize Electrical and Electronics Engineering, Bilkent University, Ankara, Turkey\\
\IEEEauthorrefmark{3}\footnotesize Asynchronous Integrated Circuits, Embedded Systems Engineering, DTU Compute, Technical University of Denmark, Denmark\\
\IEEEauthorrefmark{4}\footnotesize Asynchronous VLSI and Architecture Group, School of Engineering and Applied Science, Yale University, CT, USA\\
\IEEEauthorblockA{\footnotesize Email: \{g.leo, p.gibertini, e.covi, e.chicca\}@rug.nl, irem.ilter@ug.bilkent.edu.tr, ojuri@dtu.dk}
}



\maketitle
\begin{abstract}
Analog computing at the edge is an emerging strategy to limit data storage and transmission requirements, as well as energy consumption, and its practical implementation is in its initial stages of development. 
Translating properties of biological neurons into hardware offers a pathway towards low-power, real-time edge processing. Specifically, resonator neurons offer selectivity to specific frequencies as a potential solution for temporal signal processing.
Here, we show a fabricated \acrfull{cmos} mixed-signal \acrfull{rf} neuron circuit implementation that emulates the behavior of these neural cells responsible for controlling oscillations within the central nervous system.
We integrate the design with asynchronous handshake capabilities, perform comprehensive variability analyses, and characterize its frequency detection functionality. Our results demonstrate the feasibility of large-scale integration within neuromorphic systems, thereby advancing the exploitation of bio-inspired circuits for efficient edge temporal signal processing.
\end{abstract}
\maketitle
\input{sec1}
%
\input{sec2}
\input{sec3}

\input{sec4}
\section*{Acknowledgment}
Thanks to Muath Abu Lebdeh, Hugh Greatorex, Madison Cotteret, Willian Soares Girão, and Giacomo Indiveri for their feedback. Ole Richter and Irem Ilter performed this work while at the University of Groningen.
\bibliographystyle{IEEEtran}
\bibliography{IEEEabrv,giu_zotero,biblio}
\end{document}

%% file: sec1.tex
\section{Introduction}
\label{sec:intro}
Neurons differ in form and function, and this diversity enables the brain to produce its rich repertoire of capabilities~\cite{purves_neurosciences_2025}.
Spiking neurons can be classified according to their electrophysiological behaviour~\cite{rinzel_analysis_1998}. \Gls{lif} neurons~\cite{abbott_lapicques_1999}, predominant in modern \gls{snn} architectures, act as integrators, responding mainly to the average amplitude of the input; they belong to Class \textsc{i}, which is characterized by a continuous \gls{fi} curve.
Resonator neurons~\cite{hutcheon_resonance_2000}, whose response depends primarily on spike timing, are part of Class \textsc{ii}; they exhibit a discontinuous \gls{fi} curve, initiating high-frequency firing above a specific threshold.
They drive oscillatory activity, that controls functional states and coordinates muscular activity within the central nervous system~\cite{llinas_intrinsic_1988,wang_neurophysiological_2010}.

The \gls{rf} neuron model~\cite{izhikevich_resonate-and-fire_2001}, along with its variants~\cite{izhikevich_simple_2003,alkhamissi_deep_2021,higuchi_balanced_2024,zhang_dendritic_2025}, captures these biological behaviours compactly.
It can exhibit rhythmic firing under constant stimulation and resonance to sparse inhibitory and excitatory inputs, showing selectivity to specific frequencies, a property that can be exploited in sensory processing and temporal pattern recognition applications.
It has been successfully tested mainly in simulation across a range of applications, including associative memories~\cite{frady_robust_2019}, optical flow estimation~\cite{orchard_efficient_2021}, radar signal processing~\cite{chiavazza_low-latency_2025,reeb_range_2025,hille_resonate-and-fire_2022}, voice detection~\cite{shi_spike-vad_2025}, and gesture recognition~\cite{shaaban_resonate-and-fire_2024}. However, simulations are too computationally expensive for efficient real-time applications. To address this, digital \gls{asic} implementations have been proposed~\cite{frady_efficient_2022,orchard_efficient_2021,le_modeling_2023}. They typically rely on time-stepped numerical integration of differential equations and extensive memory updates. In contrast, analog circuits are able to exploit the inherent parallelism and continuous-time operation of the physical model.
To date, only a few analog implementations of \gls{rf} neurons have been proposed and simulated~\cite{nakada_analog_2007} or fabricated~\cite{nakada_subthreshold_2007,lehmann_direct_2023}.

\begin{figure}[b]
    \centering   
    \includegraphics[width=0.45\textwidth]{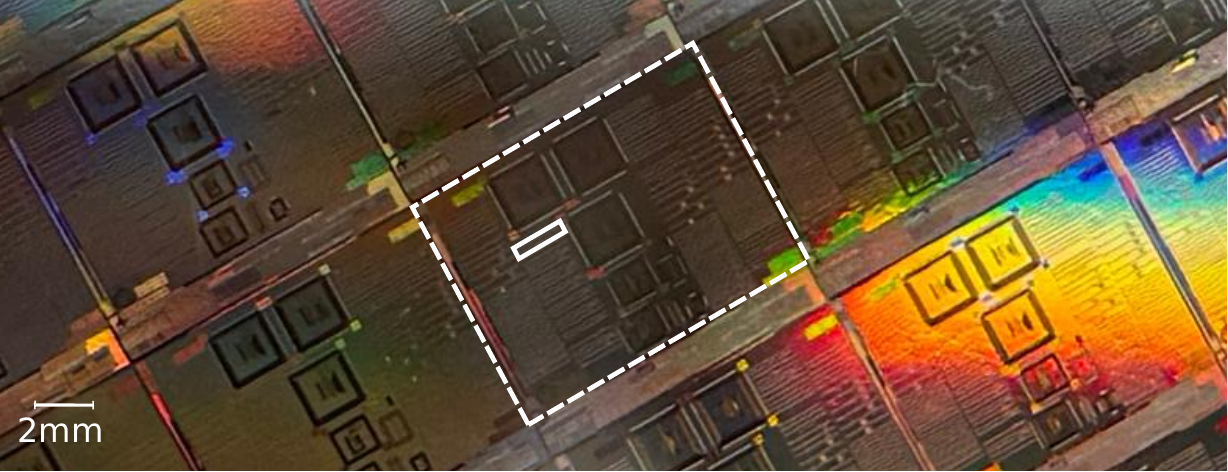} 
    \caption{Photograph of the fabricated wafer containing multiple dies. The dashed white rectangle highlights one die and the solid lined rectangle delimits the test structure that includes the \gls{rf} neuron, the I/O circuits, and the pads (\qty{122}{\micro \meter} $\times$ \qty{1772}{\micro \meter}). The neuron has an area of \qty{35}{\micro \meter} $\times$ \qty{155}{\micro \meter}.}
    \label{fig:chip}
\end{figure}

In this work, we present and characterize an enhanced \gls{cmos} implementation of the \gls{rf} neuron (Fig.~\ref{fig:chip}) originally proposed by Nakada \textit{et al.}~\cite{nakada_analog_2007}.
Our design features sub-threshold analog membrane potential dynamics and digital output spikes integrated with asynchronous handshaking for compatibility with neuromorphic hardware platforms based on \gls{aer} communication~\cite{boahen_point--point_2002}.
We fully characterized the circuit in silico across one hundred dies. Specifically, we measured variability in system parameters and overall power consumption. Furthermore, we demonstrated the selectivity properties of the spiking response with respect to the frequency of the input spike train. 
\begin{figure*}[t]
    \centering
    \includegraphics[angle=90, width=0.98\textwidth]{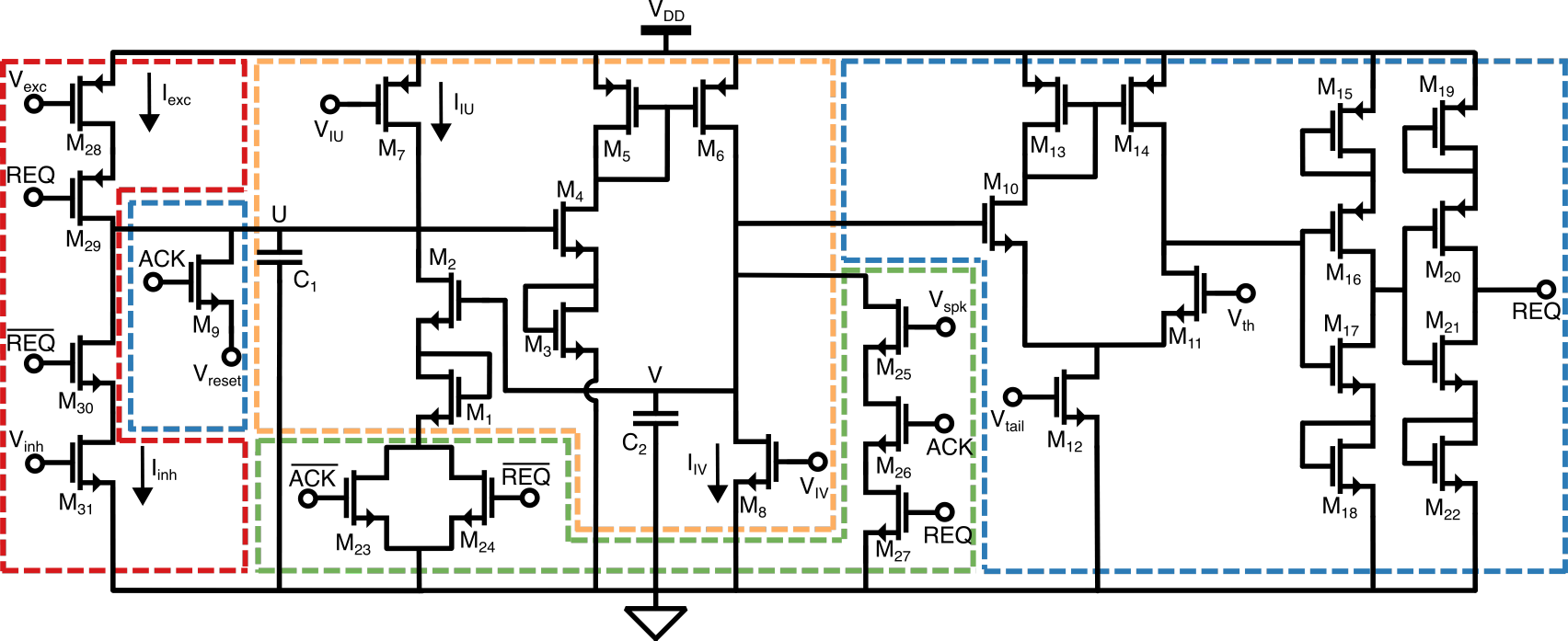}
    \caption{Schematic of the \gls{rf} circuit. The four functional blocks are highlighted in the circuit: in red, the two analog synapses, in orange, the resonator, in green, the handshake logic, and in blue, the spike generator.}
    \label{fig:schem}
\end{figure*}
We supplemented the experimental results with a broad yet low-resolution analysis of the circuit's operational range and energy consumption per spike using post-layout simulation. 
These results suggest suitability for real-time, low-power processing of audio, \gls{emg} and \gls{ecg} signals.

%

%% file: sec2.tex
%
\section{Circuit Description}
\label{sec:2}



The circuit architecture (Fig.~\ref{fig:schem}) consists of four main functional blocks: a resonator, a spike generator, a handshake logic, and two analog synapses (one inhibitory and one excitatory).

\begin{figure*}[!t]
    \centering
    \includegraphics[width=0.98\textwidth]{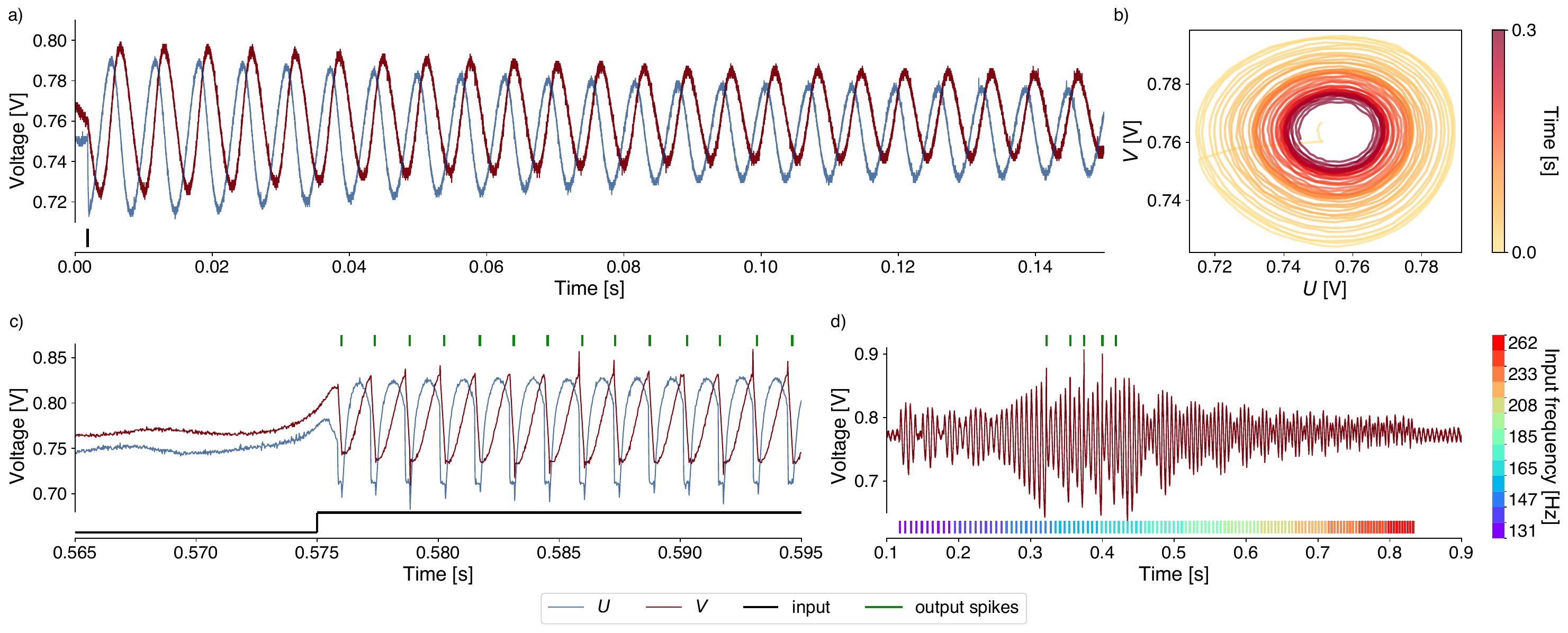}
    \caption{Measured membrane voltage dynamics for spikes and constant input. (a) Time-domain (up to \qty{0.15}{\second}) and (b) phase-plane dynamics (up to \qty{0.3}{\second}) after one inhibitory input. (c) Rhythmic firing in response to step excitatory stimulation, with a baseline value of \qty{0}{\volt} and a step value of \qty{0.5}{\volt}. (d) Response to an up-chirp consisting of 10 spikes per frequency across a total of 13 frequencies. It shows frequency-selective spiking around \qty{150}{\hertz}. The spike input in a) and d) has a time width of \qty{100}{\micro \second} and an amplitude of \qty{0.5}{\volt}. We define the inhibitory input voltage as $V_{\text{inh}}$ and the excitatory input voltage as $V_{\text{DD}} - V_{\text{exc}}$.}
    \label{fig:dynamics}
\end{figure*}
\begin{figure*}[t!]
    \centering
    \includegraphics[width=0.98\textwidth]{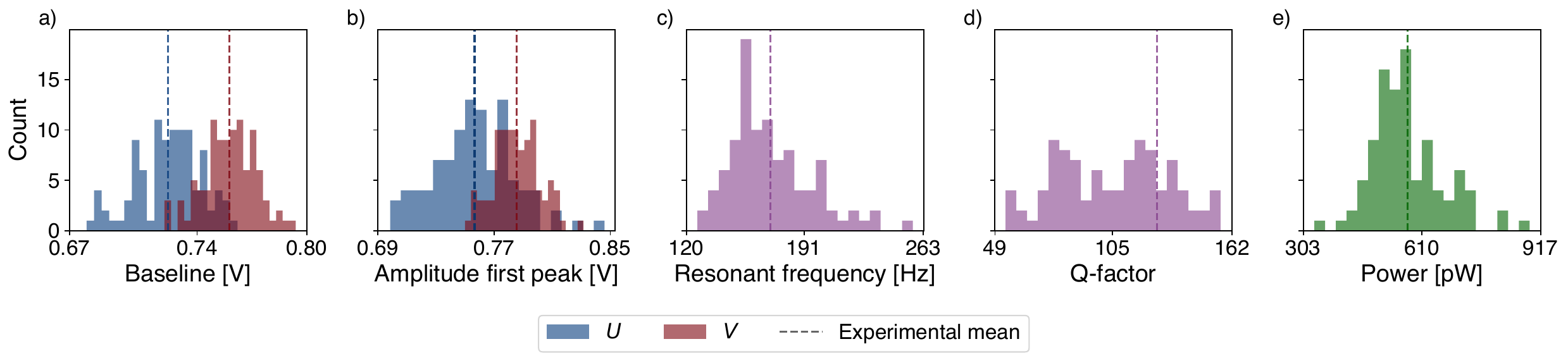}
    \caption{Measured die-to-die variability (100 dies). (a) Baseline voltage distributions for $U$ (mean value \qty{724}{\milli \volt}) and $V$ (mean value \qty{758}{\milli \volt}). (b) First peak amplitude distributions for $U$ (mean value \qty{757}{\milli \volt}) and $V$ (mean value \qty{786}{\milli \volt}). (c) Resonant frequency distributions (mean value \qty{170}{\hertz}). (d) Q-factor distributions (mean value \qty{129}). (e) Distribution of the power consumption (mean value \qty{571}{\pico \watt}).}
    \label{fig:var}
\end{figure*}
\subsubsection{Resonator Circuit ($M_1-M_8$)}
The circuit’s core oscillatory dynamics are based on~\cite{nakada_analog_2007} and are realized through a modified \gls{lv} oscillator~\cite{goel_volterra_1971}. 
A dynamical system with a stable focus equilibrium can be obtained by precisely tuning the circuit's loop gain and the currents $I_{\text{IV}}$ and $I_{\text{IU}}$ (through voltages $V_{\text{IV}}$ and $V_{\text{IU}}$). 
Taking into account the exponential dependence of the current on the gate voltage for sub-threshold circuits, the circuit dynamics can be modeled by
%
\begin{align}
C_1 \frac{dU}{dt} &= 
    I_{\text{in}} + I_{\text{IU}} 
    - I_{n0} \exp\!\left(
        \frac{\kappa_n^2}{\kappa_n + 1} \frac{V}{U_T}
      \right) \label{eq:u} \\[2mm]
C_2 \frac{dV}{dt} &= 
    I_{n0} \exp\!\left(
        \frac{\kappa_n^2}{\kappa_n + 1} \frac{U}{U_T}
      \right) 
    - I_{\text{IV}} \label{eq:v}
\end{align}
where $U$ and $V$ are the voltages across capacitors $C_1$ and $C_2$ respectively, $I_{n0}$ depends on the transistors' fabrication process and geometry, $U_{T}$ is the thermal voltage and $\kappa_{n}$ is the transistors' capacitive coupling coefficient. $I_{\text{in}}$ is the summation of the synaptic currents, $I_{\text{exc}}$ and $I_{\text{inh}}$.
The currents $I_{\text{IU}}$ and $I_{\text{IV}}$ are parameters that control the neuron's resonant frequency.
Under the change of variable $I_\alpha = I_{n0} \exp [U \kappa_n^2 / ((\kappa_n +1)U_T)]$ (and equivalent $I_\beta$ for $V$),~(\ref{eq:u}) and~(\ref{eq:v}) become
\begin{align}
\label{eq:u2}
    \frac{\mathrm{d} I_\alpha}{\mathrm{d}t} & \propto I_\alpha (I_\text{in} + I_{\text{IU}} -I_\beta)  \\[5pt]
    \label{eq:v2}
    \frac{\mathrm{d} I_\beta}{\mathrm{d}t} & \propto I_\beta (I_\alpha -I_{\text{IV}})  
\end{align}
which are the \gls{lv} equations for a 2-variable predator-prey system~\cite{goel_volterra_1971}. When expanded around the system's equilibrium point~\cite{nakada_analog_2007}, (\ref{eq:u2}) and (\ref{eq:v2}) approximate the \gls{rf} dynamics~\cite{izhikevich_resonate-and-fire_2001}, expressed by
\begin{align}
\frac{du}{dt} &= b\,u - \omega\,v \, +\, cI\label{eq:u_m}\\[5pt]
\frac{dv}{dt} &= \omega\,u + b\,v \;\label{eq:v_m}
\end{align}
where $u$ and $v$ represent the system state variables, $b<0$ is the decay factor, $\omega$ is the resonant frequency of the system, $I$ represents the total external input, and $c$ is the synaptic weight.
\subsubsection{Spiking Circuit ($M_9-M_{22}$)}
In order to detect when the voltage $V$ becomes higher than the threshold $V_{\text{th}}$ and trigger a \gls{req} signal, a simple comparator is used ($M_{10}-M_{14}$). The output is routed through a cascade of two inverters ($M_{15}-M_{22}$), which are designed to limit the current through diode connected transistors. The sharp output of the inverter chain is used as \gls{req} in the handshake protocol.
While the \gls{req} signal is high, the variable $U$ is reset to $V_{\text{reset}}$ through transistor $M_9$. 
%
\subsubsection{Handshake logic ($M_{23}-M_{27}$)}
To enable event-based communication, the circuit integrates an asynchronous handshake mechanism~\cite{sparso_introduction_2020}. After the \gls{req} signal becomes high and for the duration of the handshake, the membrane potentials $U$ and $V$ are held stable to prevent spurious dynamics. $U$ is stabilized at $V_{\text{reset}}$ and $V$ at $V_{\text{th}}$ by means of the transistors $M_{23}-M_{24}$ and $M_{25}-M_{27}$, respectively. Once an \gls{ack} signal is received, the normal dynamics resume, allowing the neuron to re-enter its oscillatory regime. The bias $V_{\text{spk}}$ modulates the duration of the spike.
\subsubsection{Synapses ($M_{28}-M_{31}$)}
The synapses (excitatory, $M_{28}-M_{29}$, and inhibitory, $M_{30}-M_{31}$) are implemented as direct transistor-based voltage-controlled current generators and allow the neuron to process both spiking inputs and continuous signals. Transistors $M_{29}$ and $M_{30}$ prevent synaptic current to flow through the circuit during the handshake.

%% file: sec3.tex
\section{Results}
\label{sec:3}
The circuit was fabricated in 180-nm \gls{cmos} technology for cost-effective prototyping, along with multiple other test structures on the \textit{chip-Olla} \gls{asic} and resulted in an area of \qty{35}{\micro \meter} $\times$ \qty{155}{\micro \meter}, mainly due to the size of the capacitors ($C_1, C_2$ = \qty{1.2}{\pico \farad}).
\begin{table}[b]
\centering
\caption{Circuit bias parameters}
\label{tab:bias}
\renewcommand{\arraystretch}{1.3}
\begin{tabular}{cccccc}
$V_\text{DD}$ & $V_{\text{th}}$ & $V_{\text{reset}}$ & $V_{\text{tail}}$ & $V_{\text{spk}}$ & $I_{\text{IU}}, I_{\text{IV}}$ \\
\midrule
\qty{1.5}{\volt} & \qty{850}{\milli \volt} & \qty{750}{\milli \volt} & \qty{300}{\milli \volt} & \qty{400}{\milli \volt} & \qty{150}{\pico \ampere}
\end{tabular}
\vspace{-1em}
\end{table}

The two analog membrane-potential signals and the digital spike output are routed off-chip through separate buffer stages. For all characterization experiments, the \gls{req} signal was externally connected to the \gls{ack} line for self-acknowledge~\cite{sparso_introduction_2020}.
The experimental characterization of the circuit was carried out with three experiments (Fig.~\ref{fig:dynamics}). In the first, we measured the circuit dynamics following the application of an inhibitory pulse. In the second, we analyzed the neuron response to a constant input to validate Class \textsc{ii} excitability. In the third, we measured the response to a spiking chirp signal, to study the neuron frequency selectivity property.
The default operating parameters and biases used for the characterization are listed in Table \ref{tab:bias}; in the second experiment, $V_{\text{th}}$ =  \qty{840}{\milli \volt}. 
The value of the bias current ($I_{\text{IU}} = I_{\text{IV}}$) is estimated by simulation and depends on the applied input bias voltages, $V_{\text{IV}}$ and $V_{\text{IU}}$.
\subsection{Internal dynamics analysis}
The first measurement (Figs.~\ref{fig:dynamics}a and~\ref{fig:dynamics}b) investigates the response of the circuit to a negative input pulse, a key analysis to understand the generation of post-inhibitory spikes, i.e., firing following inhibitory stimulation.
From these data, four system parameters were extracted — baseline voltage, first resonance peak amplitude, resonant frequency, and \gls{qf} — to assess fabrication-induced variability across multiple dies (Fig.~\ref{fig:var}).
The low variability in the baseline (\gls{cv} = 2\%) and in the amplitude of the initial oscillation (\gls{cv} = 3\%) is advantageous, as these parameters are critical for calibrating the spiking threshold bias. The resonant frequency and \gls{qf} define the system’s resonance and decay characteristics, and their variability has a significant impact on frequency detection. The resonant frequency shows a \gls{cv} of 15\%, which is expected since it is related to variability in the threshold voltage of transistors $M_2$ and $M_4$. The \gls{qf} exhibits substantial variability (\gls{cv} = 68\%), consistent with expectations due to its high sensitivity to process variations~\cite{nakada_analog_2007}.

\begin{figure}[t]
    \centering
    \includegraphics[width=0.48\textwidth]{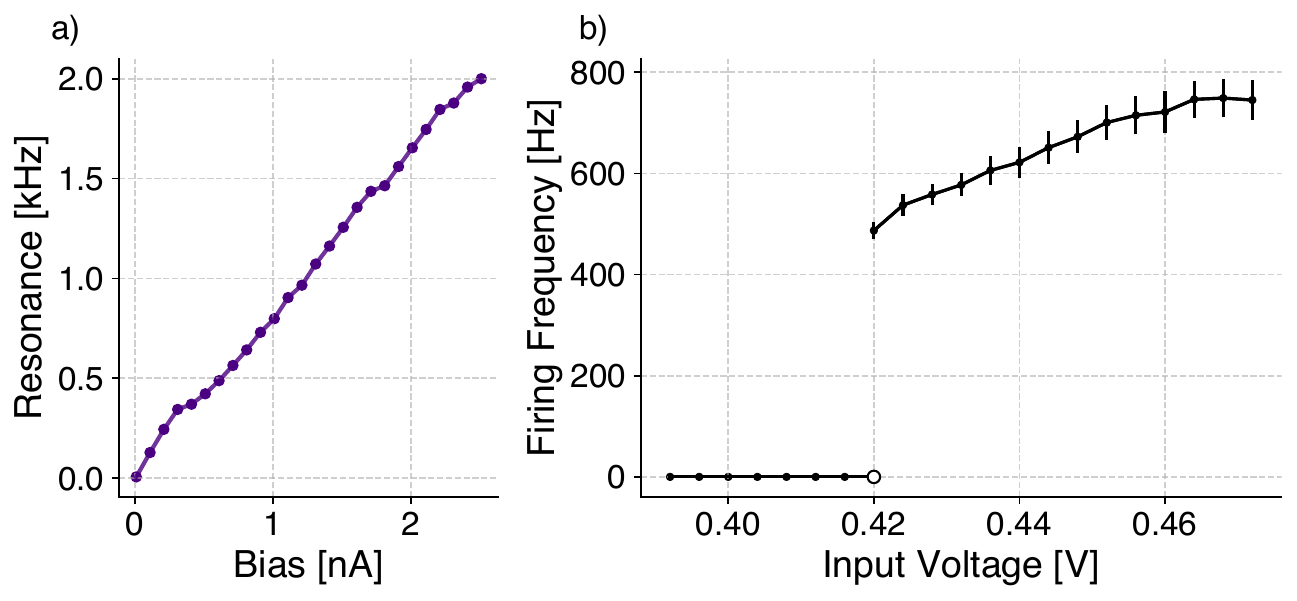}
    \caption{(a) The resonant frequency depends linearly on the bias current (post-layout simulation results), ranging from \qty{6}{\hertz} at \qty{10}{\pico \ampere} to \qty{2000}{\hertz} at \qty{2.51}{\nano \ampere}. (b) \gls{fi} curve typical of Class \textsc{ii} neurons (experimental results). The neuron is silent for inputs below \qty{420}{\milli \volt} and exhibits high-frequency rhythmic firing for larger inputs. The empty circle in the graph represents this discontinuous behavior. Each point corresponds to the mean calculated from 100 spikes, with bars showing the standard deviation.}
    \label{fig:fi_curve}
\end{figure}

We measured power consumption variability to assess feasibility of deployment in large scale systems. We observed that the power consumption during the oscillation
is approximately equal to the static power consumption;
it has a mean value of \qty{571}{\pico \watt} and a \gls{cv} of 16\%. Additionally, the estimate of the energy per spike over the entire spike duration through post-layout simulations was \qty{196}{\femto \joule}. These simulations (Fig.~\ref{fig:fi_curve}a) also indicate that the neuron’s resonant frequency, estimated through a Fourier transform analysis, is tuneable between \qtyrange{6}{2000}{\hertz} via parametrization of the circuit biases.
\subsection{Neuronal transfer function}
Experimental results (Fig.~\ref{fig:dynamics}c) for one neuron confirm that it exhibits rhythmic firing under constant input. Fig.~\ref{fig:fi_curve}b shows the \gls{fi} relationship, highlighting the Class \textsc{ii} discontinuity, after which the firing frequency scales with input.

\begin{figure}
    \centering
    \includegraphics[width=0.48\textwidth]{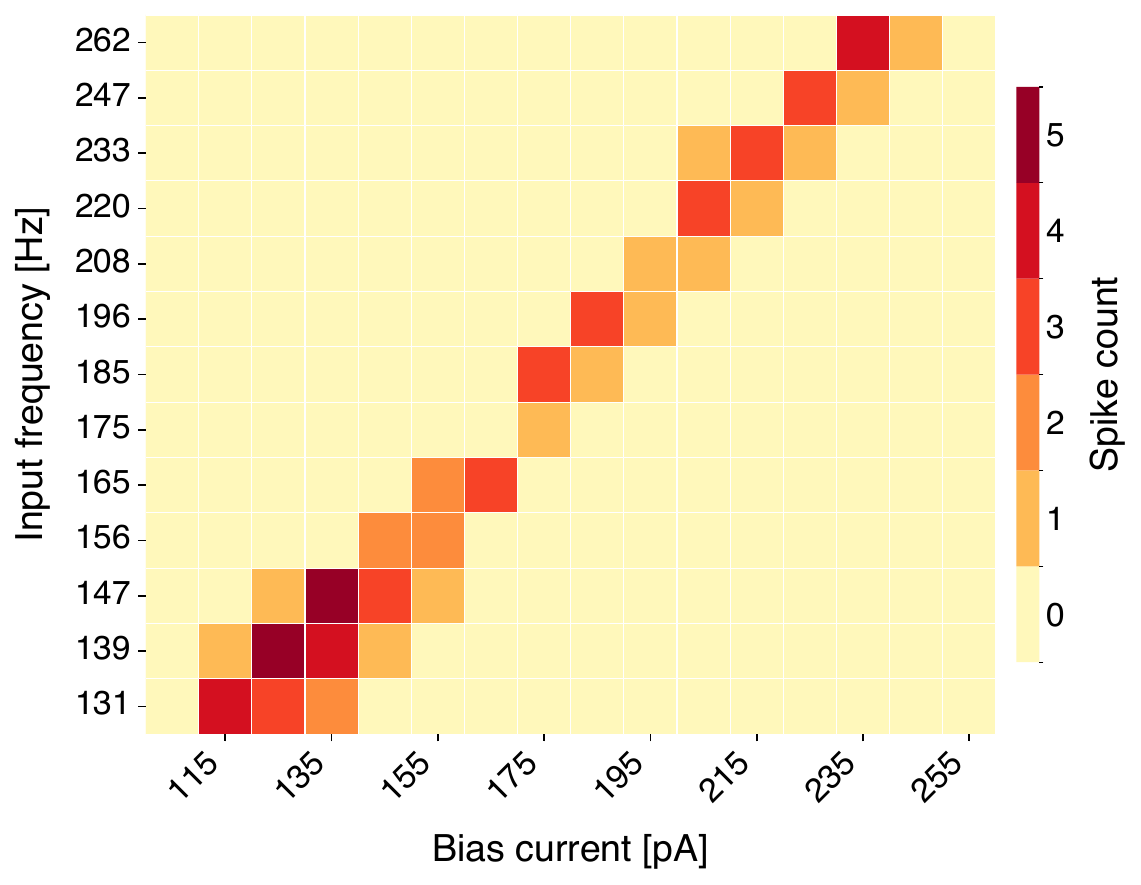}
    \caption{Measured frequency detection. The input is a chirp signal from \qty{131}{\hertz} to \qty{262}{\hertz}. The bias currents ($I_{\text{IU}} = I_{\text{IV}}$) are varied from \qtyrange{105}{255}{\pico \ampere} to tune the neuron selectivity. The threshold voltage ($V_{\text{th}}$) increases exponentially with the bias from \qtyrange{840}{900}{\milli \volt}. The results show that the neuron successfully detects specific frequencies according to the bias.}
    \label{fig:notes}
\end{figure}

Experimental measurements on the same die (Fig.~\ref{fig:dynamics}d) evaluated the neuron’s selectivity to specific input frequencies in the fine-grained range of \qtyrange{130}{260}{\hertz}. 
As shown in Fig.~\ref{fig:notes}, the circuit successfully detects specific frequencies depending on its biasing conditions. The implementation is highly flexible, as the range of detection can be adjusted via bias-current ($I_{\text{IU}}$ and $I_{\text{IV}}$) tuning, while the resolution can be modified by adjusting the spiking threshold bias $V_{\text{th}}$.

%% file: sec4.tex
\section{Conclusion}
The proposed \gls{rf} neuron implementation builds upon the circuit proposed in Nakada~\textit{et al.}~\cite{nakada_analog_2007}. Our additions include reduced power consumption, additional circuitry for embedding into \gls{aer} transceiver chips~\cite{boahen_point--point_2002}, and extensive in-silico characterization comprising die-to-die variability.

Our design exhibits low power consumption (\qty{571}{\pico \watt} versus \qty{2.34}{\micro \watt}~\cite{nakada_analog_2007}) and low energy per spike (\qty{196}{\femto \joule}, not reported in previous implementations). The study of the neuronal transfer function assessed the applicability of the system in sensory processing and band-pass filtering applications~\cite{mastella_event-driven_2024}.
Since variability is a major challenge for the large-scale deployment of analog sub-threshold neuromorphic circuits, we performed extensive variability analyses.
This proves that integration into broader neuromorphic systems~\cite{greatorex_neuromorphic_2025, richter_dynap-se2_2024} is promising; we envision combining a bank of sub-threshold, asynchronous \gls{rf} neurons with bio-inspired analog spiking sensors~\cite{bidoul_bio-inspired_2023,janotte_neuromorphic_2022} to implement efficient, event-based edge-processing systems with greater expressivity than is achievable using Class \textsc{i} neurons only~\cite{yao_spike-based_2024, narayanan_spaic_2023,schoepe_closed-loop_2023}.